\documentclass[prb,twocolumn]{revtex4-1}

\usepackage{amsmath}
\usepackage{amsfonts} 
\usepackage{graphicx} 
\usepackage{dsfont}
\usepackage{bm}

\begin{document}

\title{Generalized coherent states}

\author{T. G. Philbin}
\email{t.g.philbin@exeter.ac.uk} 
\affiliation{Physics and Astronomy Department, University of Exeter, Stocker Road, Exeter EX4 4QL, UK}
\date{\today}

\begin{abstract}
In the coherent state of the harmonic oscillator, the probability density is that of the ground state subjected to an oscillation along a classical trajectory. Senitzky and others pointed out that there are states of the harmonic oscillator corresponding to an identical oscillatory displacement of the probability density of any energy eigenstate. These generalizations of the coherent state are rarely discussed, yet they furnish an interesting set of quantum states of light that combine features of number states and coherent states. Here we give an elementary account of the quantum optics of generalized coherent states.
\end{abstract}

\maketitle 

\section{Introduction} 
The harmonic oscillator has a special significance in quantum mechanics due to its role as a bridge between the non-relativistic theory and quantum field theories. The equivalence of electromagnetic waves to a collection of non-relativistic harmonic oscillators provides the most direct route from quantum particles to quantum fields. There are many extra complications and difficulties to be encountered in quantum electrodynamics, but at the low energies relevant to quantum optics the direct connection to the harmonic oscillator is almost all that is required to begin the study of quantum light.\cite{loudon} 

Given that all quantum states of single-mode (one wave-vector, one polarization) light are states of the quantum harmonic oscillator, we can expect some important quantum states of light to have been discovered early in the twentieth century, without their significance being fully appreciated at the time. It is nevertheless impressive that what has become known as the coherent state was identified by Schr\"{o}dinger as early as 1926.\cite{sch26} It is also remarkable that a natural generalization of the coherent state to a hierarchy of similar states is rarely discussed in quantum mechanics or quantum optics. The coherent state has a Gaussian probability density whose peak follows the sinusoidal trajectory $\propto\sin(\omega t+\phi)$ of the classical particle. This leads to the description of the coherent state as a ``displaced" ground state since its probability density differs from that of the ground state only through its oscillation in time. Senitzky showed in 1954\cite{sen54} that the obvious generalization of the coherent state, in which any other energy eigenstate is displaced so that its probability density oscillates according to the classical trajectory, is also a solution for the harmonic oscillator. After brief attention in the 1950s, these displaced energy eigenstates were rediscovered in the 1970s.\cite{boi73,mar78} Roy and Singh\cite{roy82} introduced the designation ``generalized coherent states" (GCS) for Senitzky's solutions, and their properties in quantum optics were discussed by Oliveira {\it et al.}\cite{oli90}  Nieto\cite{nie97} provides a helpful history of the literature on GCS, which should be supplemented by Marhic's rediscovery,\cite{mar78} often cited\cite{fuj80,mar90,eng90,won96,kry13,mah13a,mah13b} as the original reference for these states. There have also been a few subsequent rediscoveries of GCS.\cite{yan94,lop13,bohm} So-called ``squeezed" versions of GCS, in which the probability densities pulsate in time as well as following a sinusoidal trajectory, have also been studied.\cite{ple56,sat85,mol96,nie97,kry13}

Interestingly, GCS have also been introduced independently in work on coupled field-matter systems, where they provide a useful set of basis states for some calculations. As our interest here is in the free oscillator, or single optical mode, we refer the reader to some relevant literature and the citations therein.\cite{cri92,iri05,iri07,mcc10}

GCS are as basic and elementary as the coherent state, yet they receive scant attention in comparison. They provide a facinating set of quantum states of light that combines features of two stalwarts of quantum-optics courses: number states and coherent states. GCS are a natural and instructive accompaniment to the theory of coherent states, and as such they merit entry into the textbooks of quantum optics. Indeed, GCS are simpler than squeezed states, which are standard textbook material, and they are just as interesting from the theoretical point of view. Our aim here is to set out the basic quantum optics of GCS, in a manner accessible to any student with an elementary knowledge of single-mode quantum light. Currently the best single source for the quantum optics of GCS is Oliveira {\it et al.}'s analysis,\cite{oli90} but some of the formulae there can be simplified and additional visualisations can be given. Moreover, the behaviour of GCS at a beamsplitter, which we discuss below, provides perhaps the most striking demonstration of how GCS combine properties of coherent states and number states.

\section{Generalized coherent states (GCS)} 
We consider the quantum harmonic oscillator with unit mass ($m=1$) and Hamiltonian
\begin{equation}  \label{Hho}
\hat{H}=\frac{1}{2}{\hat{p}}^2+\frac{1}{2}\omega^2{\hat{x}}^2,
\end{equation}
and throughout we set $\hbar=1$. The following wave function is a solution of the time-dependent Schr\"{o}dinger equation for this oscillator:\cite{sen54,mar78}
\begin{align}
\psi_{n,\alpha}(x,t)=& \frac{(\omega/\pi)^{\frac{1}{4}}}{2^{\frac{n}{2}}\sqrt{n!}}e^{-\frac{\omega}{2}\left(x-\langle \hat{x}\rangle\right)^2} H_n[\sqrt{\omega}(x-\langle \hat{x}\rangle)] \nonumber  \\
&\times e^{i\left[-\left(n+\frac{1}{2}\right)\omega t+x\langle \hat{p}\rangle-\frac{1}{2}\langle \hat{x}\rangle\langle \hat{p}\rangle\right]},   \label{GCS}
\end{align}
where $\langle \hat{x}\rangle$ and $\langle \hat{p}\rangle$ are the time-dependent expectation values of position and momentum in this state $\psi_{n,\alpha}(x,t)=\langle x|n,\alpha\rangle$:
\begin{align}
\langle \hat{x}\rangle=&\langle n,\alpha| \hat{x}|n,\alpha\rangle=\sqrt{\frac{2}{\omega}}|\alpha|\cos(\omega t-\theta),   \label{xav} \\
\langle \hat{p}\rangle=&\langle n,\alpha| \hat{p}|n,\alpha\rangle=-\sqrt{2\omega}|\alpha|\sin(\omega t-\theta).   \label{pav} 
\end{align}
In (\ref{GCS}), $n$ is a non-negative integer and $H_n(z)$ are the Hermite polynomials; the solution (\ref{GCS}) also depends on an arbitrary constant complex number $\alpha=|\alpha|e^{i\theta}$ that appears in (\ref{xav}) and (\ref{pav}). Note that (\ref{GCS}) is separated into an $(x,t)$-dependent amplitude and an $(x,t)$-dependent phase factor. 

\begin{figure}[h!]
\centering
\includegraphics[width=8.6cm]{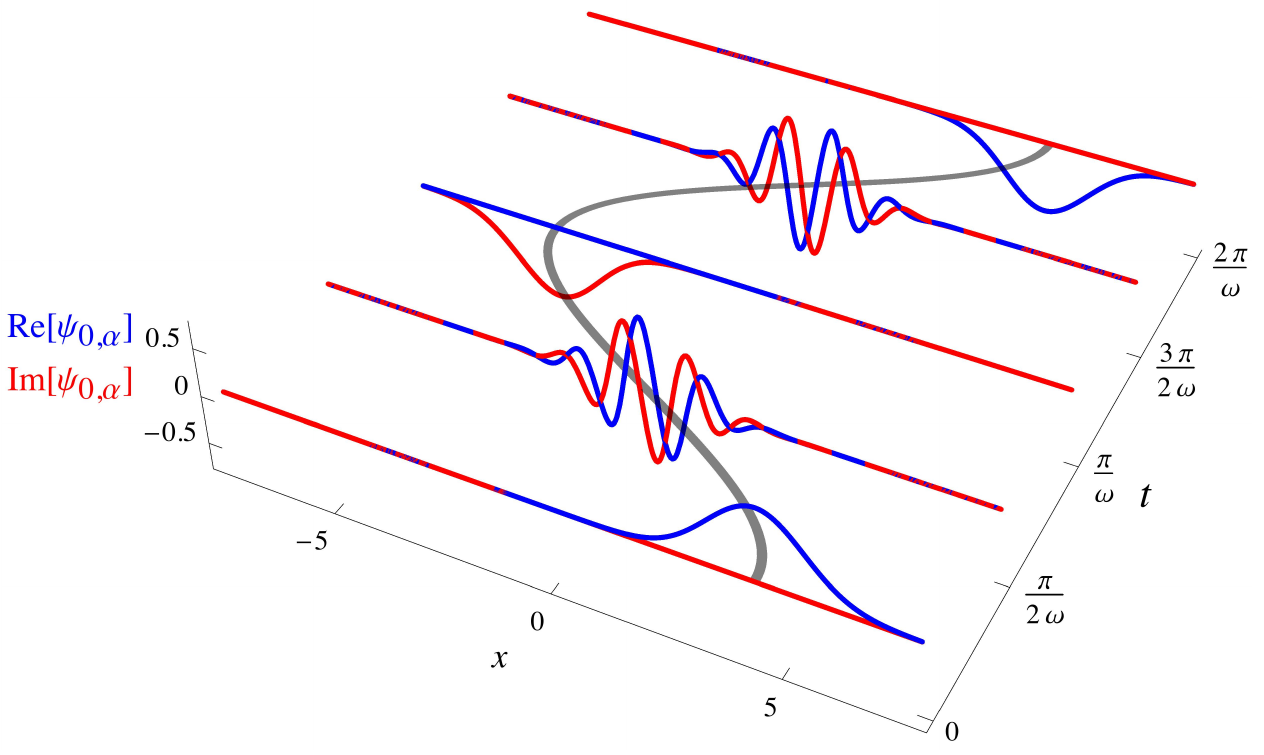}
\includegraphics[width=8.6cm]{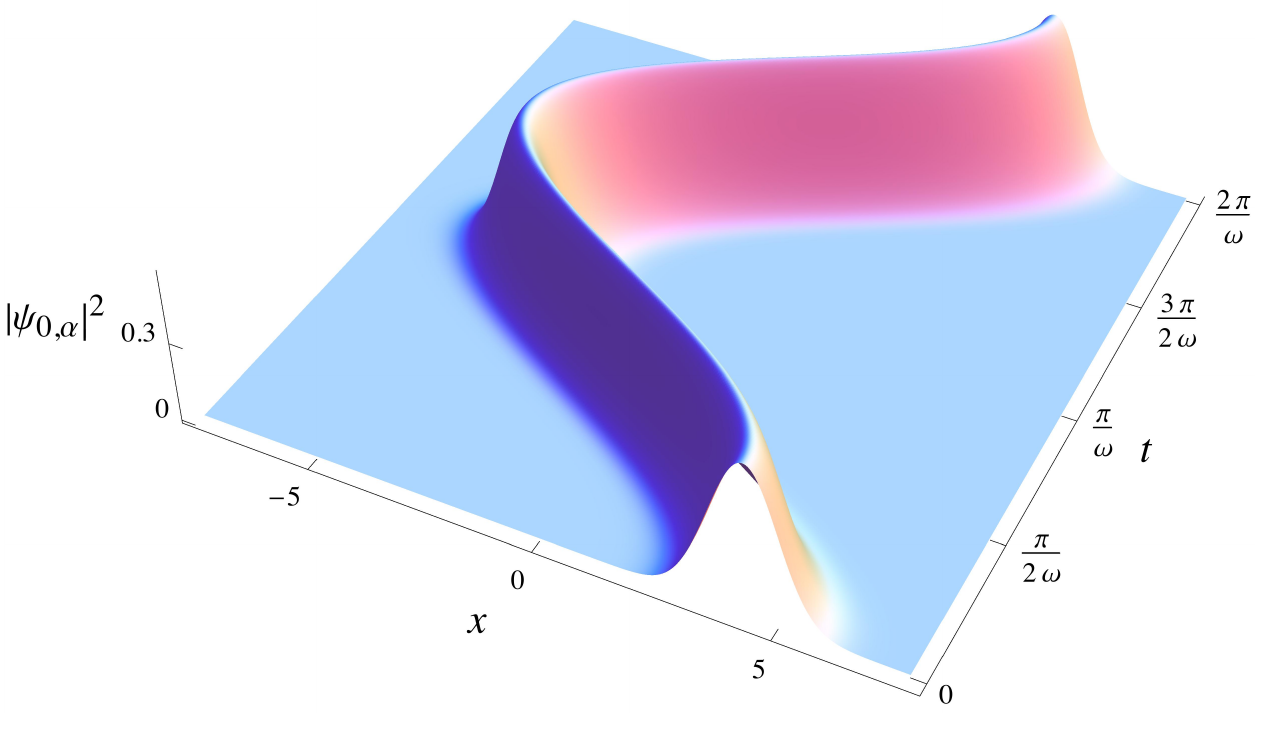}
\caption{The coherent state $\psi_{0,\alpha}(x,t)$ with $\alpha=3$ and $\omega=1$. Top: the real and imaginary parts of the wave function plotted for one period $0\leq t\leq 2\pi/\omega$. Note that the wave function acquires an overall sign change after one period $2\pi/\omega$; two periods are required for the wave function to repeat because of the zero-point energy term $\omega t/2$ in the phase. Also shown is the classical trajectory with the same amplitude of oscillation. Bottom: the probability density. }
\label{fig:gcs0}
\end{figure}

\begin{figure}[h!]
\centering
\includegraphics[width=8.6cm]{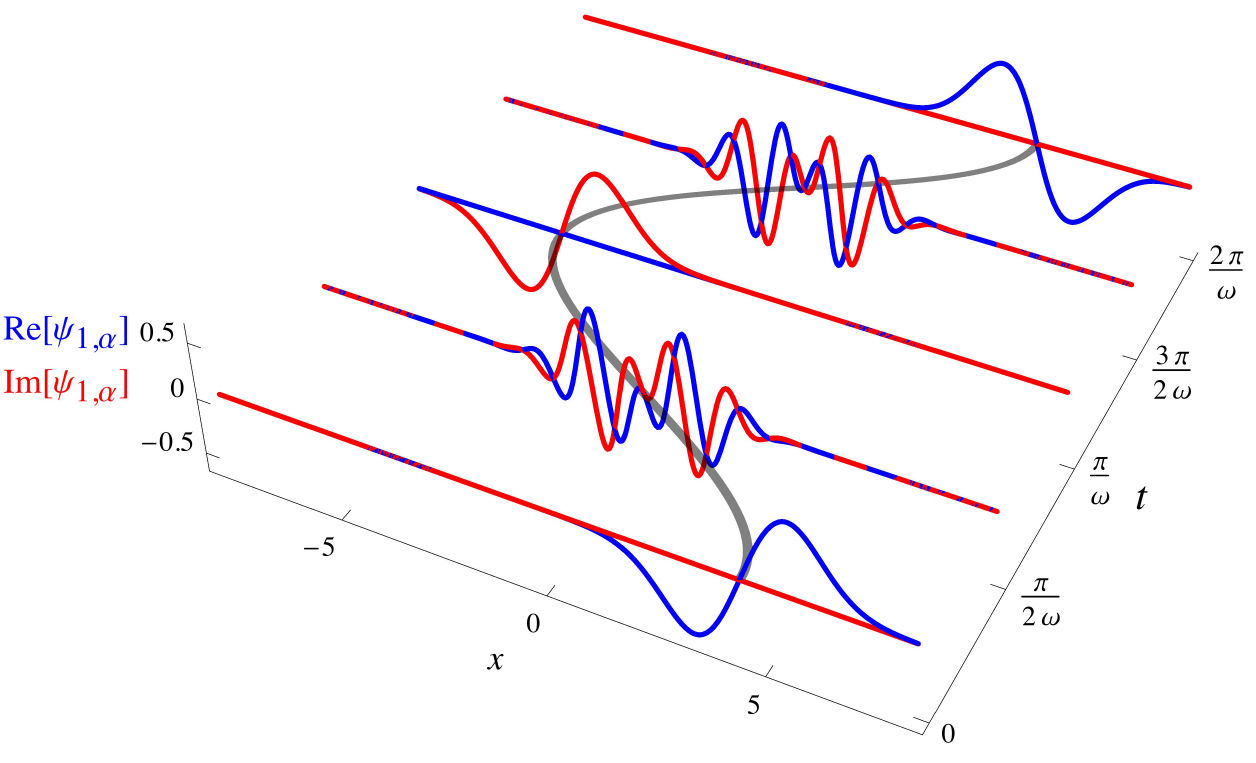}
\includegraphics[width=8.6cm]{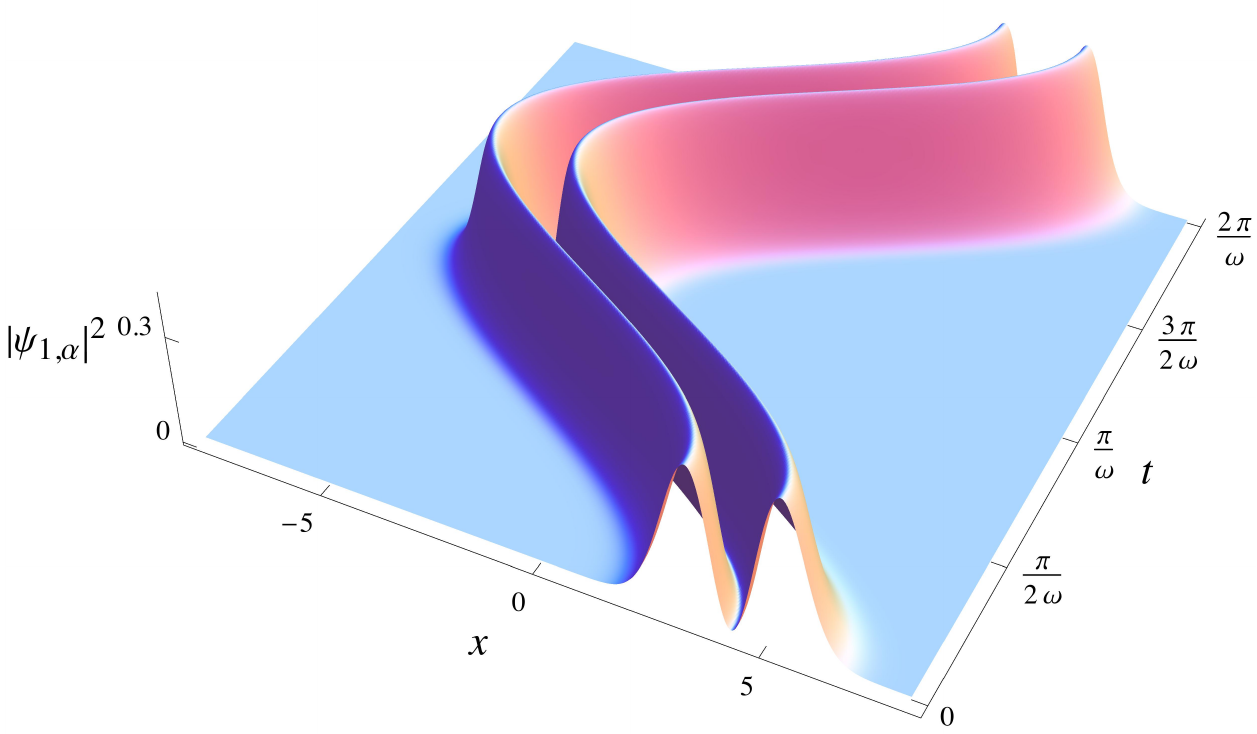}
\caption{The first generalized coherent state $\psi_{1,\alpha}(x,t)$. Parameters are the same as in Fig.~\ref{fig:gcs0}. }
\label{fig:gcs1}
\end{figure}

\begin{figure}[h!]
\centering
\includegraphics[width=8.6cm]{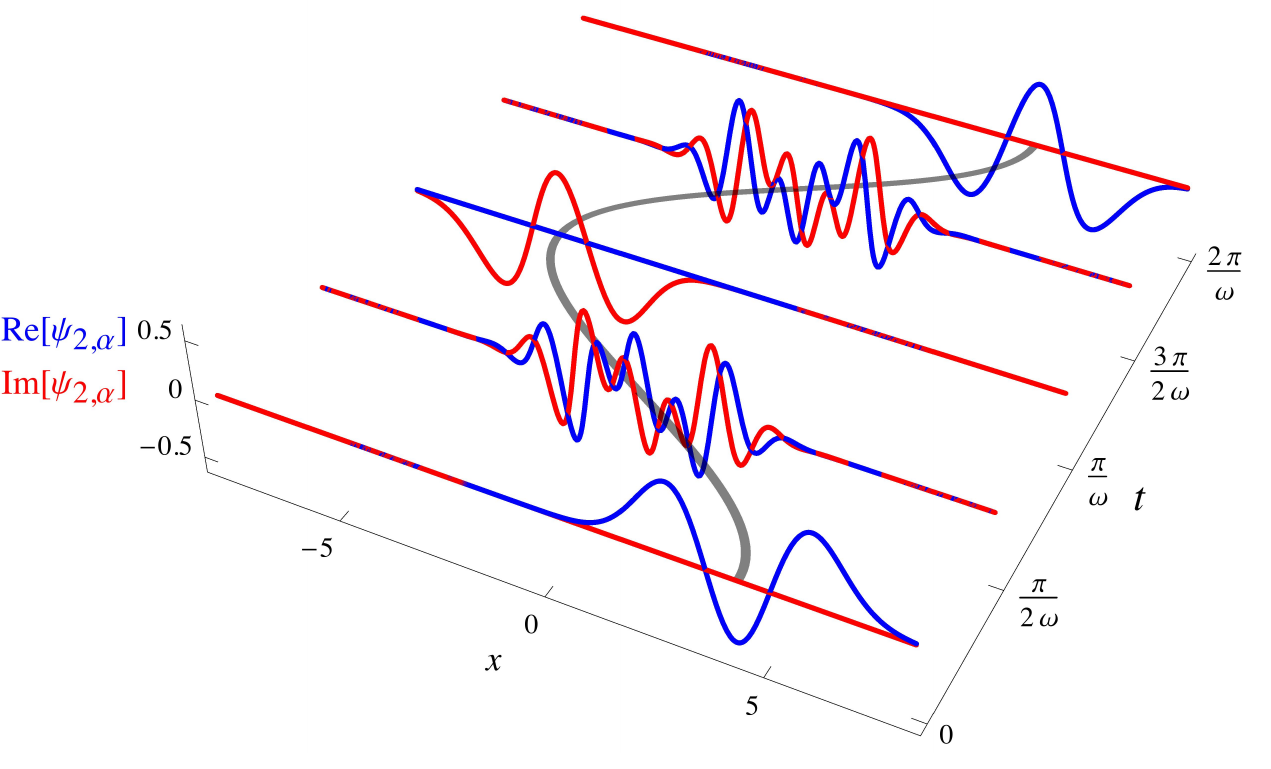}
\includegraphics[width=8.6cm]{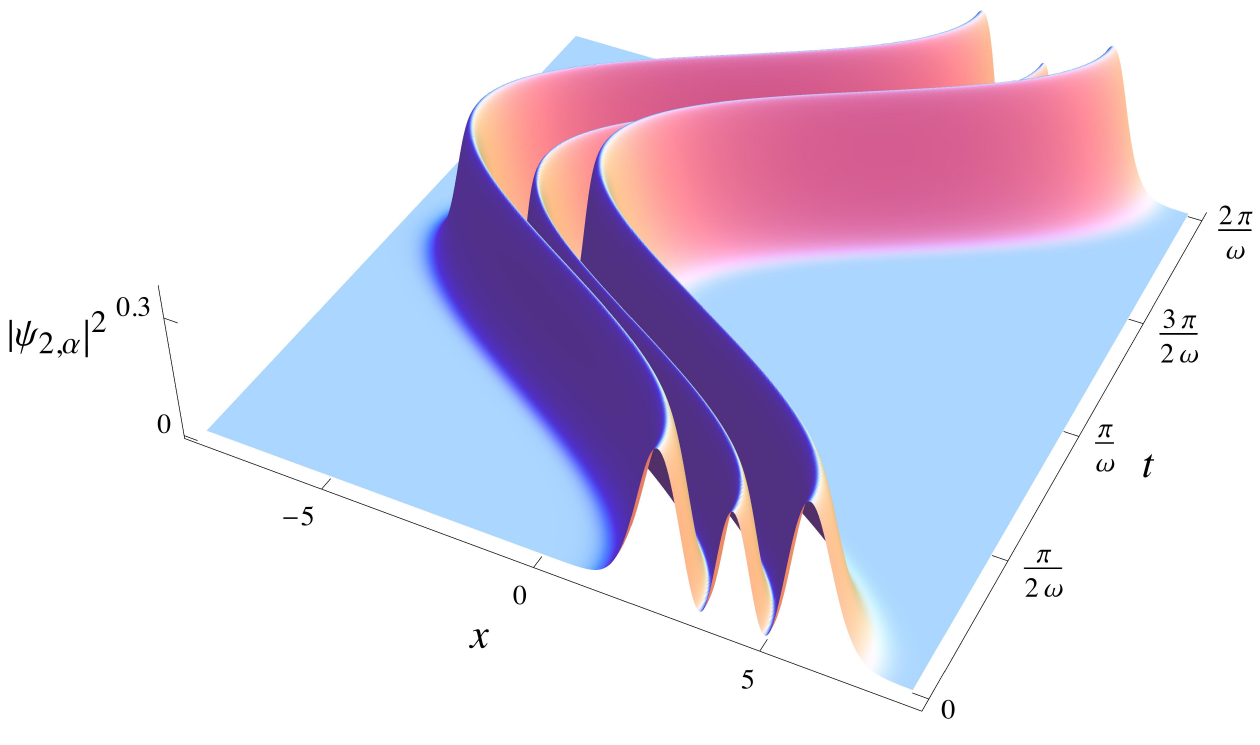}
\caption{The second generalized coherent state $\psi_{2,\alpha}(x,t)$. Parameters are the same as in Fig.~\ref{fig:gcs0}. }
\label{fig:gcs2}
\end{figure}

When $\alpha=0$, the solution (\ref{GCS}) reduces to the energy eigenstates $\psi_{n,0}(x,t)$ labelled by $n$ (including their time dependence). The set of solutions (\ref{GCS}) thus differs from the energy eigenstates in the following manner: the $x$-dependence of the amplitude is displaced by the time-dependent term (\ref{xav}) and the phase is displaced by an amount $x\langle \hat{p}\rangle-\frac{1}{2}\langle \hat{x}\rangle\langle \hat{p}\rangle$. The probability density of (\ref{GCS}) (the square of the amplitude) therefore maintains exactly the shape of the static density $|\psi_{n,0}(x,t)|^2$, but performs an oscillation in time that follows a classical trajectory. The dynamics of the states $\psi_{n,\alpha}(x,t)$ are depicted in Figs.~\ref{fig:gcs0}--\ref{fig:gcs2} for $n=0,1,\ \text{and}\ 2$.

For $n=0$, the solution (\ref{GCS}) is the coherent state. Sch\"{o}dinger's discovery of $\psi_{0,\alpha}(x,t)$ was motivated\cite{sch26} by the desire to find a quantum state that would reproduce the classical motion at macroscopic scales.\cite{cp} For the coherent state, the Gaussian spread in $x$ of the probability density becomes smaller as a fraction of the amplitude $|\alpha|\sqrt{2/\omega}$ of its oscillation as $|\alpha|$ increases. The classical motion at macroscopic energies with trajectory $x(t)=|\alpha|\sqrt{2/\omega}\cos(\omega t-\theta)$ can then be viewed as an approximation to a coherent state with large $|\alpha|$. This picture is validated in quantum optics by results that show the coherent state to be the closest state to a classical plane wave allowed by quantum mechanics.\cite{loudon} Note that the same argument concerning the spread of the probability density as a fraction of the amplitude of its oscillation can be made for any GCS, since they all feature the same Gaussian factor $e^{-\omega\left(x-\langle \hat{x}\rangle\right)^2/2}$. But the states $\psi_{n,\alpha}(x,t)$ for $n>0$ are \emph{not} classical-like states. This will become clear in the quantum-optics setting of GCS, through their measurement properties and behaviour at a beamsplitter. In fact, the most interesting aspect of GCS is how they combine a ``quantum number" $\alpha$ that imparts classical features to the state with a quantum number $n$ that imparts quantum features. 

\begin{figure}[h!]
\centering
\includegraphics[width=7.99cm]{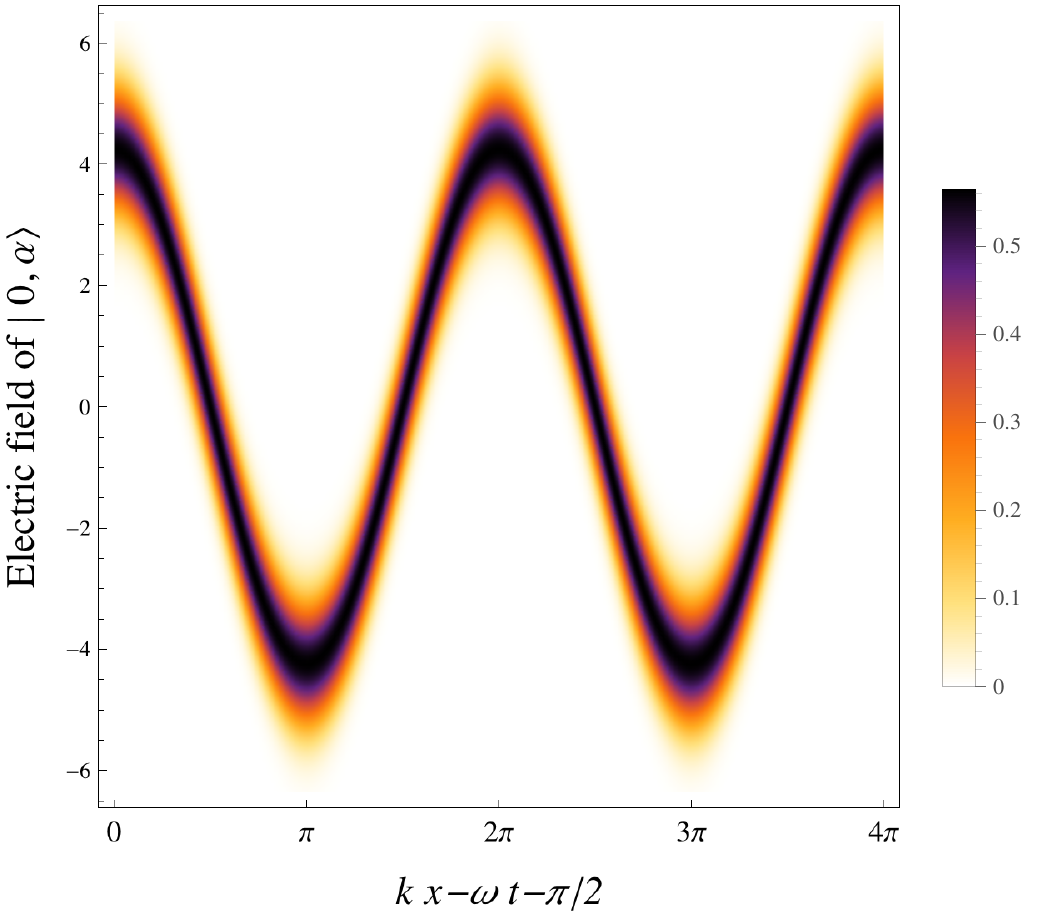}
\includegraphics[width=7.99cm]{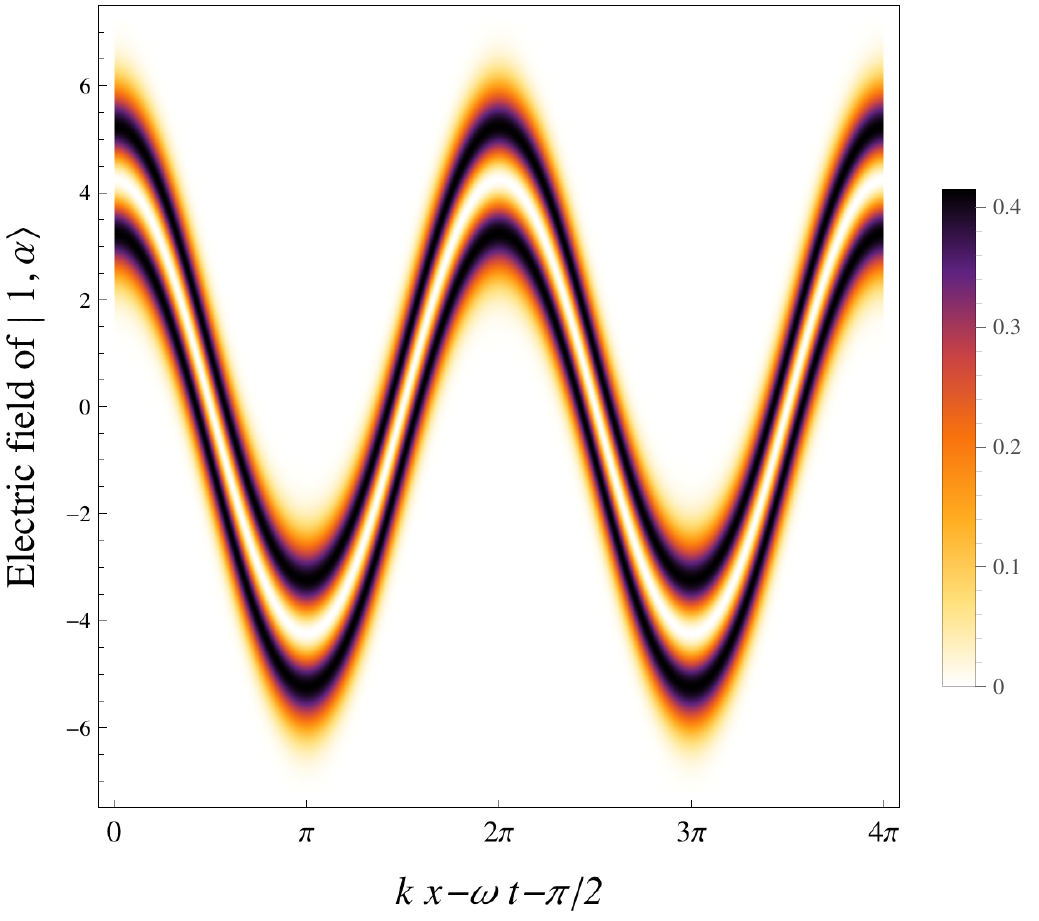}
\includegraphics[width=7.99cm]{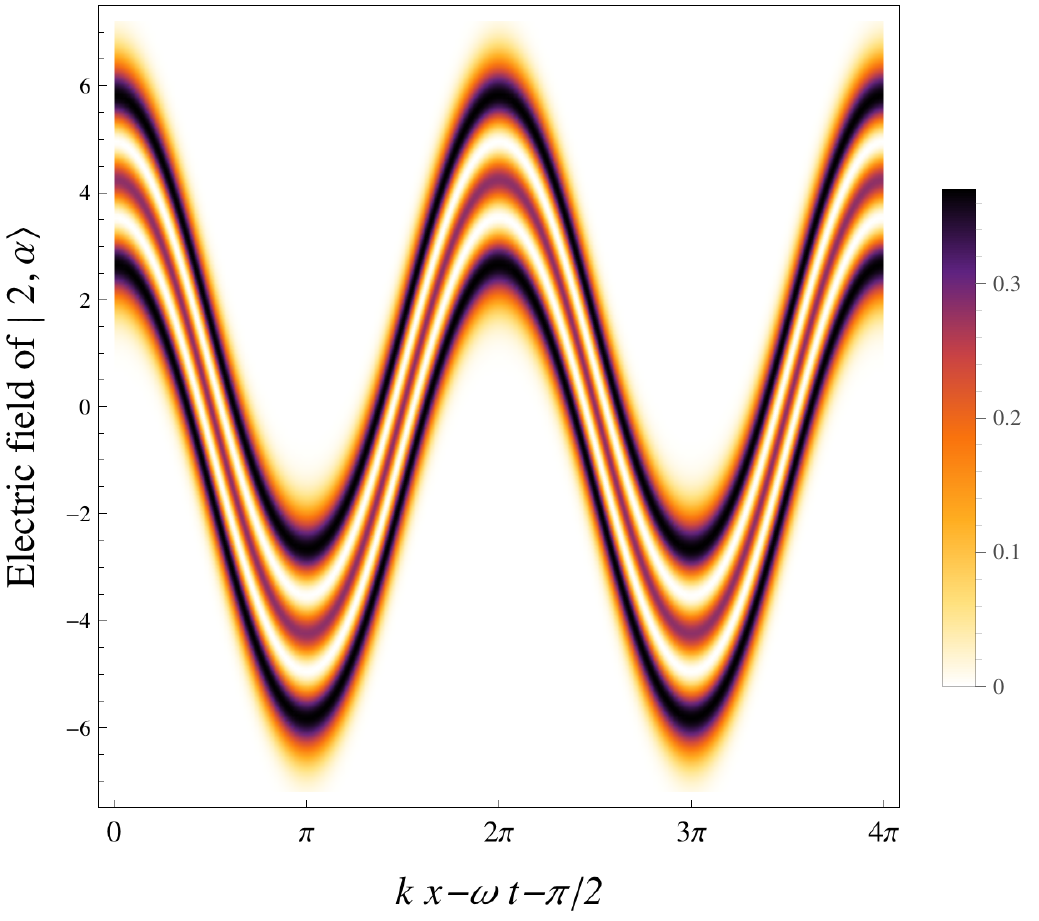}
\caption{Electric-field probability distributions in $|n,\alpha\rangle$ for $n=0,\ 1$ and $2$, with $\alpha=3$ and $\omega=c=1$. Darker colours represent higher probabilities.}
\label{fig:gcse}
\end{figure}

The position and momentum operators for the oscillator are related to the creation and annihilation operators by
\begin{equation}  \label{xaad}
\hat{x}=\frac{1}{\sqrt{2\omega}}\left(\hat{a}+\hat{a}^\dagger\right), \quad \hat{p}=-i\sqrt{\frac{\omega}{2}}\left(\hat{a}-\hat{a}^\dagger\right).
\end{equation}
Here we are using the Schr\"{o}dinger picture and have included in (\ref{GCS}) the time dependence of the energy eigenstates $|n,0\rangle$, so that we have
\begin{align}  
\hat{a}|n,0\rangle=&\sqrt{n}\,e^{-i\omega t}|n-1,0\rangle,    \label{anum1} \\
 \hat{a}^\dagger|n,0\rangle=&\sqrt{n+1}\,e^{i\omega t}|n+1,0\rangle.  \label{anum2}
\end{align}
The coherent state $|0,\alpha\rangle$ is an eigenstate of $\hat{a}$ ($=\sqrt{\omega/2}(x+\omega^{-1}d/dx)$ in the coordinate representation (\ref{GCS})):
\begin{equation}   \label{a0al}
\hat{a}|0,\alpha\rangle=\alpha e^{-i\omega t}|0,\alpha\rangle,
\end{equation}
an equation usually written for $t=0$. The transition to single-mode quantum optics\cite{loudon} is achieved through replacement of the position $\hat{x}$ by the electric-field operator
\begin{equation}  \label{Eaad}
\hat{E}(x)=\frac{1}{\sqrt{2\omega}}\left(\hat{a}e^{ikx+\pi/2}+\hat{a}^\dagger e^{-ikx-\pi/2}\right), \quad k=\omega/c,
\end{equation}
which is time-independent in the Schr\"{o}dinger picture. The GCS $|n,\alpha\rangle$ then have electric-field probability distributions that are equal to the probability distributions for the position of the oscillator but with $\omega t$ replaced by $\omega t-kx-\pi/2$. In Fig.~\ref{fig:gcse} we plot the electric-field probability distributions for the first three GCS, which correspond to the position probability distributions in  Figs.~\ref{fig:gcs0}--\ref{fig:gcs2}. The electric field of the coherent state $|0,\alpha\rangle$ has the familiar appearance of a plane wave with a noise band. For general GCS $|n,\alpha\rangle$, the electric field shows $n+1$ noise bands separated by $n$ nodes. In the photon number states $|n,0\rangle$ these noise bands do not oscillate, i.e.\ they show no dependence on the phase $kx-\omega t+\pi/2$. Many of the properties of the number states $|n,0\rangle$ are preserved by GCS $|n,\alpha\rangle$ with $\alpha>0$, as we shall see. The number of nodes in the electric-field probability distribution is thus the important signature of these properties, not the number of photons in the state; the latter is uncertain for GCS and can be of arbitrarily large average value for any $n$.

\section{GCS as a basis}
Using the representation (\ref{GCS}) one can verify that the orthonormality relation $\langle n,0|m,0\rangle=\delta_{nm}$ for number states is preserved for GCS with the same complex amplitude $\alpha$:
\begin{equation}
\langle n,\alpha|m,\alpha\rangle=\delta_{nm}.
\end{equation}
For given $n$, however, GCS with different complex amplitudes are not orthogonal, as is familiar for the coherent state $|0,\alpha\rangle$; the general relation is
\begin{equation}
\langle n,\beta|n,\alpha\rangle=e^{-(|\alpha|^2+|\beta|^2-2\alpha\beta^*)/2}\,L_n\left(|\alpha-\beta|^2\right),
\end{equation}
where $L_n(z)$ are the Laguerre polynomials. The overlap between two GCS with equal $n$ thus becomes exponentially small as the difference $|\alpha-\beta|$ in their amplitudes increases, just as for coherent states.

Ladder operators exist for the quantum number $n$ of $|n,\alpha\rangle$ and are in fact given by displaced versions of the number-state ladder operators $\hat{a}$ and $\hat{a}^\dagger$. Defining
\begin{equation}  \label{aal}
\hat{a}_\alpha=\hat{a}-\alpha e^{-i\omega t} \quad \Longrightarrow \quad \left[\hat{a}_\alpha,\hat{a}_\alpha^\dagger\right]=1,
\end{equation}
one can show using the representation (\ref{GCS}) that
\begin{align}  
\hat{a}_\alpha|n,\alpha\rangle=&\sqrt{n}\,e^{-i\omega t}|n-1,\alpha\rangle,    \label{agcs1} \\
 \hat{a}^\dagger_\alpha|n,\alpha\rangle=&\sqrt{n+1}\,e^{i\omega t}|n+1,\alpha\rangle,  \label{agcs2}
\end{align}
which are the generalizations of (\ref{anum1}) and (\ref{anum2}). The displacement (\ref{aal}) of $\hat{a}$ is effected by the displacement operator\cite{loudon}
\begin{gather}
\hat{D}_\alpha=\exp\left(\alpha e^{-i\omega t}\hat{a}^\dagger-\alpha^* e^{i\omega t}\hat{a}\right),  \label{Ddef}  \\
\hat{D}_{-\alpha}^\dagger  \hat{a}   \hat{D}_{-\alpha} =\hat{a}-\alpha e^{-i\omega t} =\hat{a}_\alpha,   \label{Dal}
\end{gather}
and GCS are displaced number states:
\begin{equation}  \label{Dgcs}
|n,\alpha\rangle=\hat{D}_\alpha |n,0\rangle,
\end{equation}
as is verified by showing that (\ref{Dgcs}) and (\ref{Dal}) imply (\ref{agcs1}) and (\ref{agcs2}).

It should not be surprising that $|n,\alpha\rangle$ for fixed $\alpha$ form a complete basis for single-mode states (just like number states), whereas $|n,\alpha\rangle$ for fixed $n$ form an over-complete basis (just like coherent states). The relevant relations are
\begin{align}  
\sum_{n=0}^\infty |n,\alpha\rangle \langle n,\alpha| =I,    \label{comn} \\
\frac{1}{\pi}\int d^2\alpha \, |n,\alpha\rangle \langle n,\alpha| =I .  \label{comal}
\end{align}
The completeness relation (\ref{comn}) follows immediately from (\ref{Dgcs}), but the over-completeness relation (\ref{comal}) is perhaps most easily verified using the number state expansion of  $|n,\alpha\rangle$ below (equation (\ref{gcsnum})).

\section{Expectation values and uncertainty relations}
The results of the previous section give the electric-field expectation value and uncertainty in GCS:
\begin{gather}  
 \langle n,\alpha| \hat{E}(x) |n,\alpha\rangle =\sqrt{\frac{2}{\omega}}\,|\alpha| \cos\left(kx-\omega t+\theta+\frac{\pi}{2}\right),   \label{Eexp} \\
\left(\Delta E(x,t)\right)^2=\frac{2n+1}{2\omega},  \label{deltaE}
\end{gather}
where $\alpha=|\alpha|e^{i\theta}$ as before. Note from (\ref{deltaE}) that the uncertainty in the electric field is identical to that in the related number state (it is independent of $\alpha$). Note that the expectation value (\ref{Eexp}) and uncertainty (\ref{deltaE}) of the electric field does not convey the nodal structure of the electric-field probability distributions of GCS, as depicted in Fig.~\ref{fig:gcse}. The quadrature operators $\hat{X}$ and $\hat{Y}$, defined by
\begin{equation}  \label{quad}
\hat{E}(x)=\frac{2}{\sqrt{\omega}}\left[\hat{X}\cos(kx+\pi/2)+\hat{Y}\sin(kx+\pi/2)\right],
\end{equation}
have uncertainties in GCS that are also identical to those in the related number state, as is easily verified.

For the photon number operator $\hat{N}=\hat{a}^\dagger\hat{a}$ we obtain
\begin{gather}  
 \langle n,\alpha| \hat{N} |n,\alpha\rangle =n+|\alpha|^2,   \label{Nexp} \\
\left(\Delta N\right)^2=|\alpha|^2=\langle n,\alpha| \hat{N} |n,\alpha\rangle -n,  \label{deltaN}            \\
\frac{\Delta N}{\langle n,\alpha| \hat{N} |n,\alpha\rangle}=\frac{|\alpha|}{n+|\alpha|^2}.  \label{fracdeltaN}
\end{gather}
Contrary to the electric field, we see from (\ref{deltaN}) that the uncertainty in photon number is independent of $n$ and is thus identical to that in a coherent state with the same complex amplitude $\alpha$. The relation between $\Delta N$ and the expectation value of $\hat{N}$ is, however, different from the coherent state for $n>0$, as is shown by (\ref{fracdeltaN}). 

Equation (\ref{deltaN}) shows the important property of coherent states ($n=0$) that $\left(\Delta N\right)^2$ is equal to $\langle 0,\alpha| \hat{N} |0,\alpha\rangle$. This property that the variance is equal to average value is a characteristic of the Poisson distribution, and the coherent state has indeed a Poissonian photon-number distribution\cite{loudon} (the photon-number distributions for general GCS are given in the next section). Equation (\ref{deltaN}) also shows that $\left(\Delta N\right)^2$ is less than $\langle n,\alpha| \hat{N} |n,\alpha\rangle$ for $n>0$, which means that GCS for $n>0$ exhibit sub-Poissonian fluctuations,\cite{loudon} i.e the fluctuations in photon number in measurements of the states are less than those of a Poisson distribution and thus less than those of a coherent state. The photon-number probability distributions for GCS are given below and indeed differ greatly from the Poisson distribution when $n>0$. Sub-Poissonian fluctuations are a signature of nonclassical light.\cite{loudon} It is not surprising that we find by this criterion that GCS for $n>0$ are nonclassical; their nonclassical nature was already clear in Fig.~\ref{fig:gcse}. Note from (\ref{fracdeltaN}) that for all GCS the fractional uncertainty in photon number goes to zero as the amplitude $|\alpha|$ goes to infinity; the fluctuations approach Poissonian behaviour in this limit as we then have $n+|\alpha|^2 \approx |\alpha|^2$ .

\section{Number-state expansion and photon statistics}
GCS can be generated from the coherent state by using the creation operator $\hat{a}^\dagger_\alpha$ (see (\ref{agcs2})), or alternatively they can be generated from number states by using the displacement operator $\hat{D}_\alpha$ (see (\ref{Dgcs})). Either procedure allows the number-state expansion of $|n,\alpha\rangle$ to be calculated. It not a trivial matter to express the result in terms of special functions but one can verify that the number-state expansion is
\begin{align}
|n,\alpha\rangle= & \frac{1}{\sqrt{n!}} e^{-|\alpha|^2/2}   \nonumber   \\
& \times \sum_{k=0}^\infty (-1)^{n+k} \sqrt{k!}(\alpha^*)^{n-k} L_k^{n-k}\left(|\alpha|^2\right) |k,0\rangle,    \label{gcsnum}
\end{align}
where $L_k^{m}(z)$ are generalized Laguerre polynomials. From (\ref{gcsnum}) we see that the probability $P_k(n,\alpha)$ of finding $k$ photons in the GCS $|n,\alpha\rangle$ is given by
\begin{equation} \label{pnk}
P_k(n,\alpha)= \frac{k!}{n!}e^{-|\alpha|^2} |\alpha|^{2(n-k)}\left[ L_k^{n-k}\left(|\alpha|^2\right) \right]^2.
\end{equation}
For the coherent state $n=0$, equation (\ref{pnk}) is of course a Poisson distribution and $P_k(1,\alpha)$ can also be written in a simple form:
\begin{align}
P_k(0,\alpha)=& e^{-|\alpha|^2} \frac{|\alpha|^{2k}}{k!},  \\
 P_k(1,\alpha)=&e^{-|\alpha|^2} \frac{|\alpha|^{2(k-1)}(|\alpha|^2-k)^2}{k!}.
\end{align}

\begin{figure}[h!]
\centering
\includegraphics[width=8.6cm]{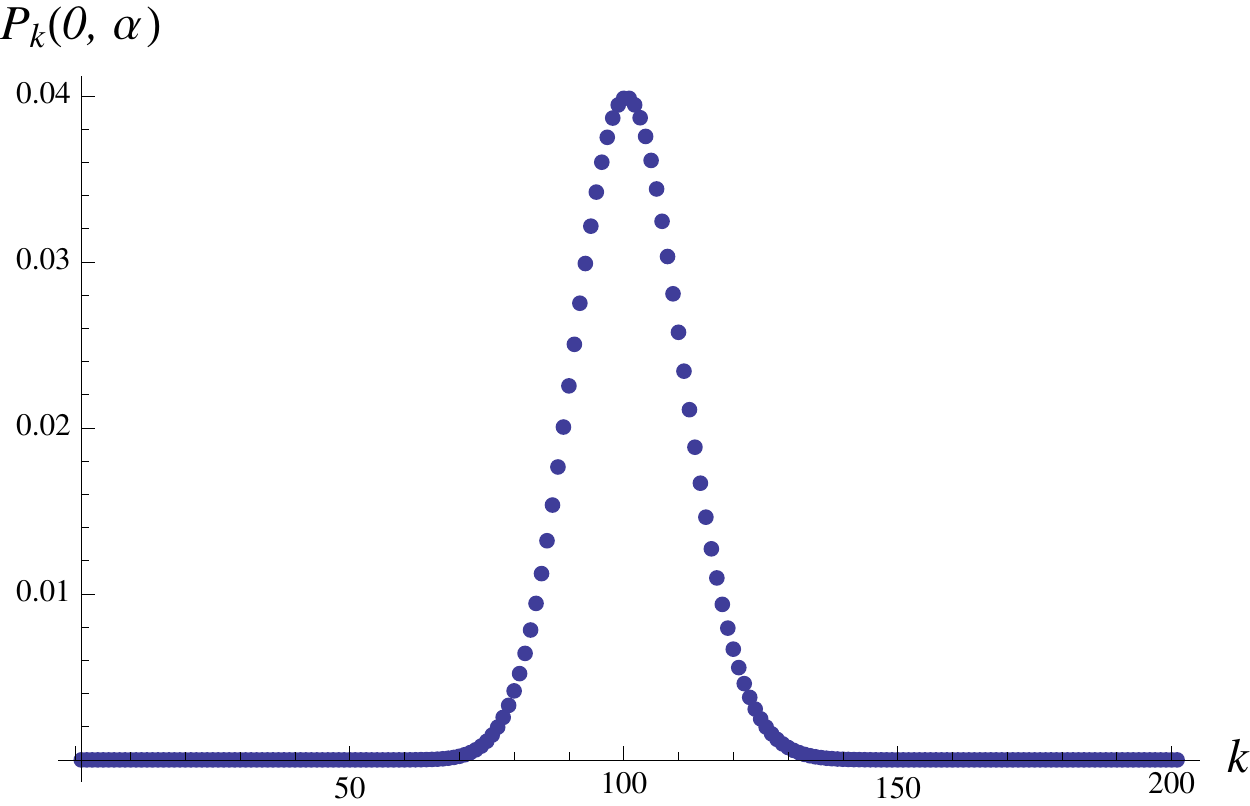}
\includegraphics[width=8.6cm]{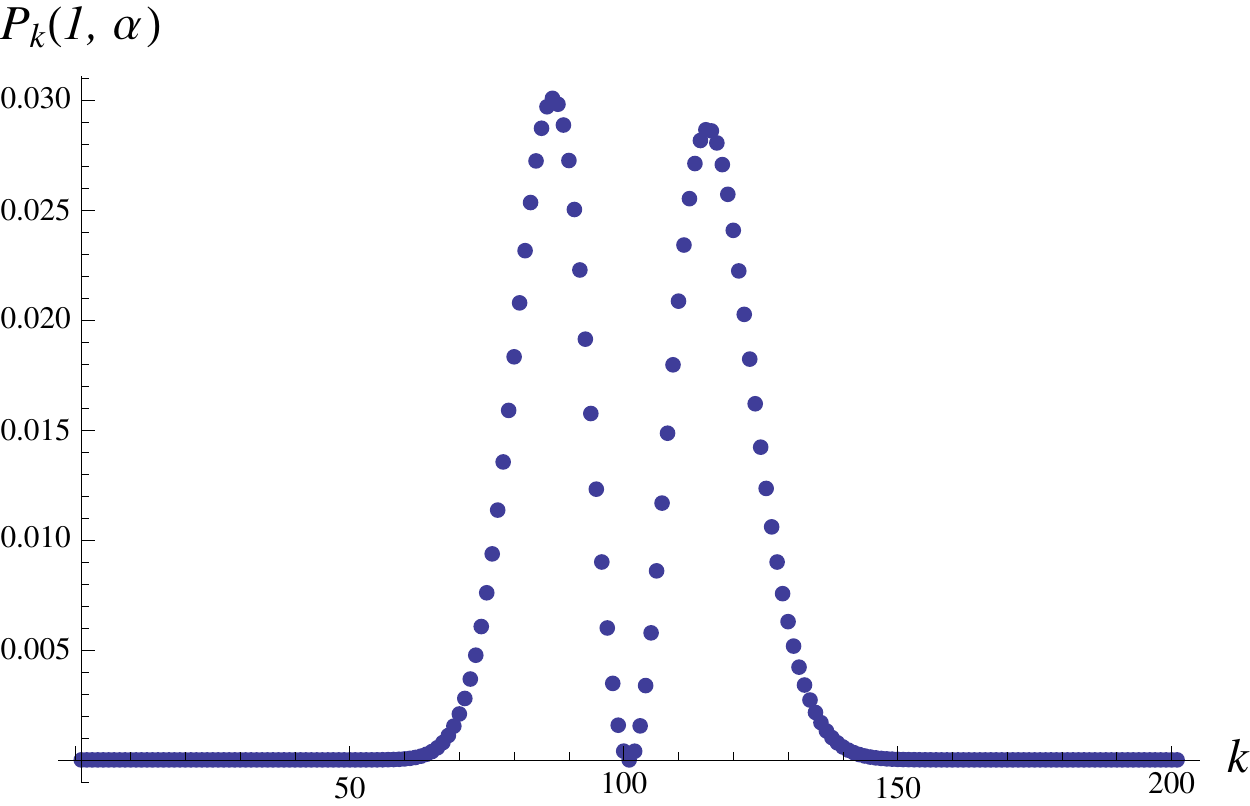}
\includegraphics[width=8.6cm]{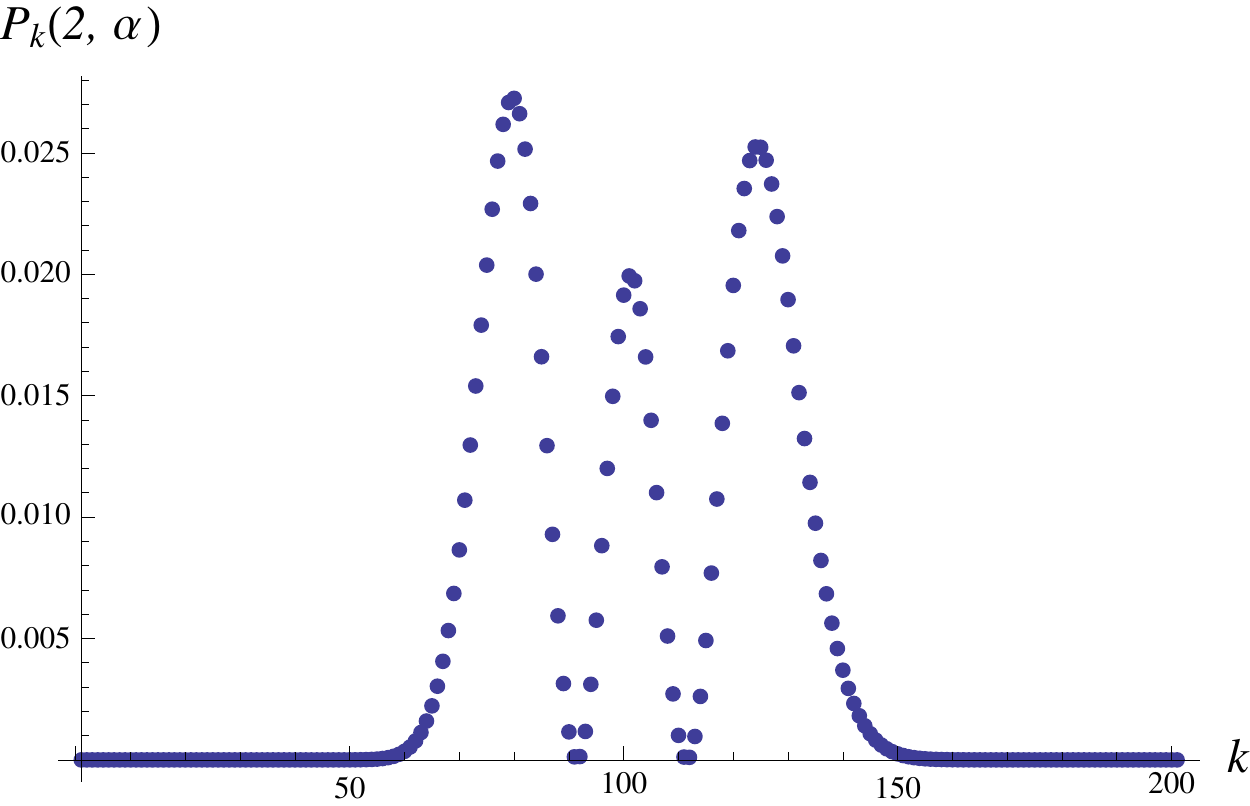}
\caption{Photon probability distributions in $|n,\alpha\rangle$ for $n=0,\ 1$ and $2$, with $|\alpha|=10$. }
\label{fig:pkn}
\end{figure}

The distributions $P_k(n,\alpha)$ are plotted in Fig.~\ref{fig:pkn} for $n=0,\ 1$ and $2$. The most striking feature of these distributions is a similarity of their general shapes to those of the electric-field probability densities (and thus to the probability densities of the harmonic-oscillator energy eigenstates). But note that that there are in general no nodes in the distributions $P_k(n,\alpha)$, just local minima, and the relative sizes of the local maxima in $P_k(n,\alpha)$  are also not the same as in the electric-field probability densities. Oliveira {\it et al.}\cite{oli90} have given a phase-space interpretation of the oscillations in the photon-number distributions.

\section{Second-order coherence}
Optical coherence is an extremely complicated subject\cite{mandel} but basic measures of coherence are not difficult to define and calculate. First-order coherence is essentially a measure of correlations of the electric field in the beam, while second-order coherence is a measure of intensity correlations. The correlation function of the electric field, the average $\langle E((x_1,t_1) E((x_2,t_2)\rangle$ of the product of the electric field at two space-time points, becomes in quantum optics an expectation value quadratic in the annihilation and creation operators (see (\ref{Eaad})). Similarly, in quantum optics the intensity correlation function $\langle I((x_1,t_1) I((x_2,t_2)\rangle$ takes the form of an expectation value quartic in $\hat{a}$ and $\hat{a}^\dagger$. The ordering of $\hat{a}$ and $\hat{a}^\dagger$ in these expectation values is chosen to correspond to expressions for measured intensities.\cite{loudon} When the resulting expectation values are normalized, so that they take the value 1 when all space-time points coincide, they form the quantum measures of first- and second-order coherence, respectively. The resulting degree of first-order coherence, denoted by $g^{(1)}(x_1,t_1;x_2,t_2)$, simplifies greatly for a plane parallel single-mode beam such as a GCS. In fact, for such beams it is always the case that $|g^{(1)}(x_1,t_1;x_2,t_2)|=1$, and the beam is said to be first-order coherent.\cite{loudon} The second-order coherence function, denoted $g^{(2)}(x_1,t_1;x_2,t_2)$, takes the following simple form for a plane parallel single-mode beam\cite{loudon} 
\begin{equation}
g^{(2)}(x_1,t_1;x_2,t_2)=\frac{\langle \hat{a}^\dagger\hat{a}^\dagger\hat{a}\hat{a}\rangle}{\langle \hat{a}^\dagger\hat{a}\rangle^2}.
\end{equation}
This expression is easily rewritten in terms of the photon number operator and can then be evaluated for GCS using (\ref{Nexp}) and (\ref{deltaN}), as follows:
\begin{align}
g^{(2)}(x_1,t_1;x_2,t_2)&=\frac{\langle \hat{N}^2 \rangle-\langle \hat{N} \rangle}{\langle \hat{N} \rangle^2}=1+\frac{(\Delta N)^2-\langle \hat{N} \rangle}{\langle \hat{N} \rangle^2}    \nonumber  \\
&=1-\frac{n}{(n+|\alpha|^2)^2}    \label{g2}
\end{align}
A beam is said to be second-order coherent if $g^{(2)}=1$ in addition to $|g^{(1)}|=1$, which for GCS occurs only for the coherent state $n=0$. For $n>0$, GCS have $g^{(2)}<1$, which is another signature of nonclassical light.\cite{loudon} In the limit $n\to\infty$, or in the limit $|\alpha|\to\infty$, GCS approach second-order coherence 
$g^{(2)}\to1$. 

For given $n$, $g^{(2)}$ for GCS is minimized by the number states $\alpha=0$. Interestingly, for given $\alpha>0$ the minimum in $g^{(2)}$ is at the value of $n$ closest to $|\alpha|^2$; from (\ref{Nexp}) we see that this corresponds to an equal contribution from $n$ and $\alpha$ to the average photon number in the state. If  $|\alpha|^2$ is an integer this minimum in $g^{(2)}$ for fixed $\alpha>0$ is $g^{(2)}=1-1/(4|\alpha|^2)$. 

\section{Behaviour at a beamsplitter}
\begin{figure}[h!]
\centering
\includegraphics[width=4cm]{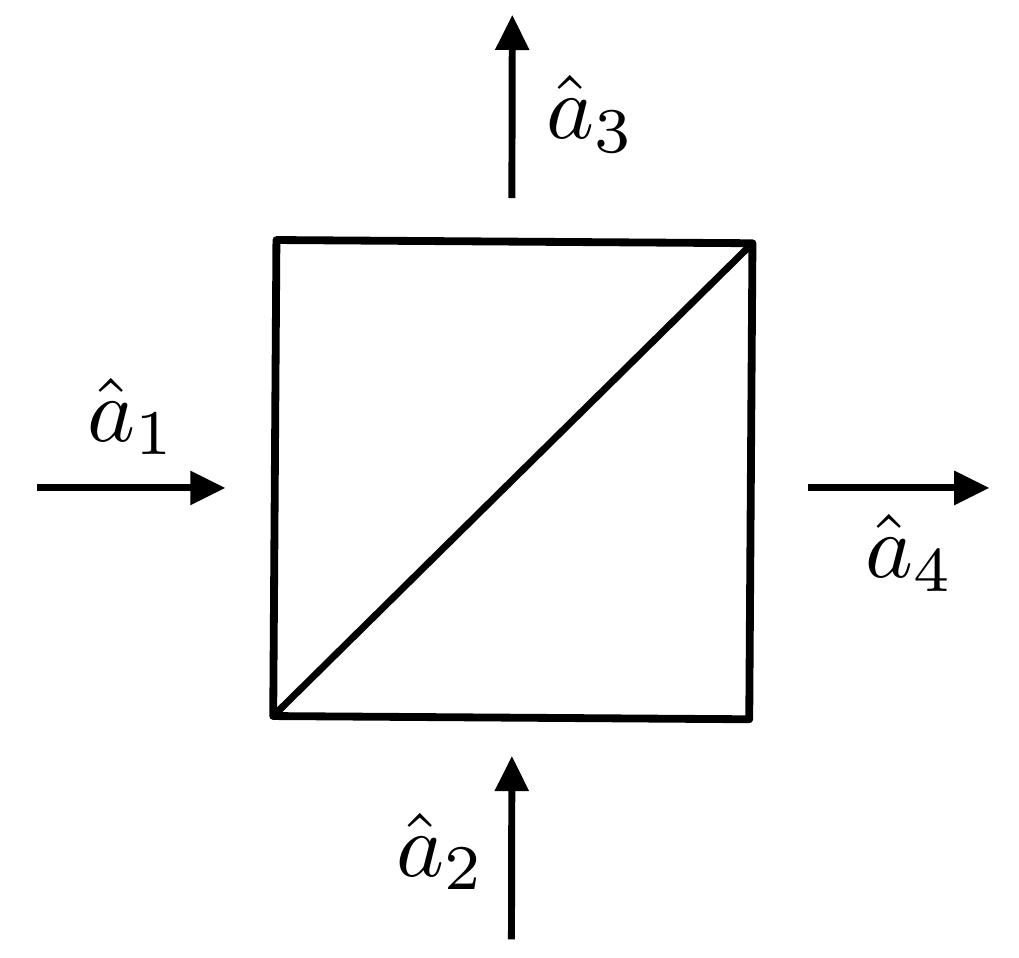}
\caption{Beamsplitter with input and output arms showing the annihilation operators for the corresponding modes.}
\label{fig:bs}
\end{figure}

The results so far have shown that GCS combine features of coherent states and number states. This is perhaps most vividly illustrated by the behaviour of the GCS $|n,\alpha\rangle$ at a beamsplitter, where we will find that the ``$\alpha$" part of the state behaves like a coherent state and the ``$n$" part behaves like a number state. We consider a symmetric beam splitter with input arms 1 and 2, and output arms 3 and 4 (see Fig.~\ref{fig:bs}). The beamsplitter input-output relations are given by\cite{loudon}
\begin{gather}
\hat{a}_3=R\hat{a}_1+T\hat{a}_2, \quad \hat{a}_4=T\hat{a}_1+R\hat{a}_2,   \\
\hat{a}_1=R^*\hat{a}_3+T^*\hat{a}_4, \quad \hat{a}_2=T^*\hat{a}_3+R^*\hat{a}_4, \label{a134}  \\
|R|^2+|T|^2=1,  \quad RT^*+TR^*=0,
\end{gather}
where $R$ and $T$ are the complex reflection and transmission coefficients. These are the relations that would be satisfied classically by the complex amplitudes of the electric field in the arms of the beamsplitter.\cite{loudon}

Consider a GCS entering the beam splitter in arm 1, with arm 2 in the vacuum state. The input state is thus $|n,\alpha\rangle_1\, |0,0\rangle_2$, which can be written
\begin{gather}
|n,\alpha\rangle_1\, |0,0\rangle_2=\hat{D}_{1\alpha}\frac{1}{\sqrt{n!}}\left(e^{-i\omega t}\hat{a}_1^\dagger\right)^n|0,0\rangle_1\, |0,0\rangle_2,   \label{12input}  \\
\hat{D}_{1\alpha}=\exp\left(\alpha e^{-i\omega t}\hat{a}^\dagger_1-\alpha^* e^{i\omega t}\hat{a}_1\right). \label{D1al}
\end{gather}
In (\ref{12input}) the state $|n,\alpha\rangle_1$ is built up from the number state $|n,0\rangle_1$, created using $n$ factors of $\hat{a}_1^\dagger$, followed by the displacement operator $\hat{D}_{1\alpha}$ for arm 1, which creates the GCS from the number state as in  (\ref{Dgcs}). Recall that we include the time dependence in the number states and in GCS. By substitution for $\hat{a}_1^\dagger$ using (\ref{a134}) the number state $|n,0\rangle_1$ is expressed in terms of the output modes in the standard manner:\cite{loudon}
\begin{gather}
|n,0\rangle_1=\frac{1}{\sqrt{n!}}\left(e^{-i\omega t}\hat{a}_1^\dagger\right)^n|0,0\rangle_1   \nonumber   \\
=\sum_{m=0}^n\left(\frac{n!}{m!(n-m)!}\right)^{1/2}R^mT^{n-m}|m,0\rangle_3\,|n-m,0\rangle_4.   \label{numinput}
\end{gather}
It is also a standard result\cite{loudon} that the displacement operator (\ref{D1al}) for arm 1 becomes a product of displacement operators in the output arms when (\ref{a134}) is employed: $\hat{D}_{1\alpha}=\hat{D}_{3R\alpha}\hat{D}_{4T\alpha}$. Using this last result and (\ref{numinput}) in (\ref{12input}) gives
\begin{align}
|n,\alpha\rangle_1\, |0,0\rangle_2 =& \sum_{m=0}^n\left(\frac{n!}{m!(n-m)!}\right)^{1/2}R^mT^{n-m}  \nonumber   \\
&\times |m,R\alpha\rangle_3\, |n-m,T\alpha\rangle_4.   \label{gcsout}
\end{align}
For $\alpha=0$, the right-hand side of (\ref{gcsout}) gives the familiar superposition of number states in the output arms of the beamsplitter, produced when a number-state is sent into one arm. For $n=0$, the result (\ref{gcsout}) is the familiar product state of independent coherent states in the output arms, produced by a coherent-state input. For general GCS with nonzero $\alpha$ and $n$, we have in (\ref{gcsout}) the remarkable result that the quantum number $n$ of GCS behaves at a beamsplitter exactly like a number state with $n$ photons, while the ``quantum number" $\alpha$ behaves exactly like a coherent state with complex amplitude $\alpha$. The quantum number $n$ of GCS in no way measures the number of photons in the state; instead it measures the number of nodes in the electric-field probability distribution, and the number of local minima in the photon distribution. Yet the $n$ of GCS produces the same entanglement at a beamsplitter as that produced by $n$ photons. Each electric-field node in arm 1 has a probability amplitude $R$ to be reflected into arm 3 and a  probability amplitude $T$ to be transmitted into arm 4, just like a photon. For the case of $n=1$ we have from (\ref{gcsout})
\begin{align}
|1,\alpha\rangle_1\, |0,0\rangle_2 = R|1,R\alpha\rangle_3\, |0,T\alpha\rangle_4+T|0,R\alpha\rangle_3\, |1,T\alpha\rangle_4.   \label{0alout}
\end{align}
Thus, if the GCS in the middle plot of Fig.~\ref{fig:gcse} is sent through a beam splitter and both output arms are measured, then one output arm will contain the same state as the input, but with reduced amplitude, while the other output arm will contain a coherent state.
Results like (\ref{0alout}) are particularly interesting because the input GCS in this relation can have arbitrarily large (average) energy, yet it produces the same entanglement at the beamsplitter as a single photon. 

\section{Generation of GCS}
Returning to the case of the quantum harmonic oscillator, let us recall the exact solution for dynamically generating Schr\"{o}dinger's coherent state from an initial ground state.\cite{merzbacher} The coherent state is generated from the ground state by acting on it with the displacement operator (equation (\ref{Dgcs}) with $n=0$). The time-development operator $\hat{T}(t,t_0)$ takes the state $|\psi(t_0)\rangle$ at time $t_0$ to the state $|\psi(t)\rangle$ at time $t$: $|\psi(t)\rangle=\hat{T}(t,t_0)|\psi(t_0)\rangle$. It follows that to generate the coherent state from the ground state, the time-development operator must be equal to the displacement operator, up to phase factors. If the oscillator is subjected to a classical external force $f(t)$, which corresponds to adding a term $f(t)\hat{x}$ to the Hamiltonian (\ref{Hho}), then the dynamics can be solved exactly and the resulting time-development operator is\cite{merzbacher}
\begin{align}
\hat{T}(t,t_0)=&e^{i\beta(t,t_0)}\exp\left[\zeta(t,t_0)\hat{a}^\dagger e^{-i\omega t}-\zeta^*(t,t_0)\hat{a}e^{i\omega t} \right]      \nonumber   \\
&\times e^{-i \hat{H}_0(t-t_0)/\hbar},   \label{Top}
\end{align}
where $\hat{H}_0$ is the Hamiltonian without the external driving term $f(t)\hat{x}$ (i.e. the  Hamiltonian (\ref{Hho})), $\zeta(t,t_0)$ is given by
\begin{equation}
\zeta(t,t_0)=-\frac{i}{\sqrt{2\omega}}\int_{t_0}^tdt'\,f(t')e^{i\omega t'},
\end{equation}
and the c-number real phase $\beta(t,t_0)$ is given by
\begin{equation}
\beta(t,t_0)=\frac{1}{2\omega}\int_{t_0}^tdt'\int_{t_0}^{t'} dt''\,f(t')f(t'')\sin[\omega(t'-t'')].
\end{equation}
If the initial state $|\psi(t_0)\rangle$ is an energy eigenstate, then (\ref{Top}) shows that the driving force gives a time-development operator that is a displacement operator (\ref{Ddef}), up to a c-number phase factor. Thus the oscillator is driven into a coherent state by a classical external force if $|\psi(t_0)\rangle$ is the ground state, an important and well-known result in quantum mechanics.\cite{merzbacher} From (\ref{Dgcs}) we also find that a classical external force will drive the oscillator into the GCS $|n,\alpha\rangle$, with $\alpha$ determined by the external force and the interaction time, if the initial state is the energy eigenstate $|n,0\rangle$.

The foregoing results apply also to single-mode quantum light interacting with a classical external current $\bm{j}$, since the coupling term in the Hamiltonian is $\bm{j\cdot\hat{A}}$, where $\bm{\hat{A}}$ is the vector potential operator.\cite{merzbacher} There are differences in the formulae due to the spatial dependence of the quantum fields and the fact that the current is coupled to $\bm{\hat{A}}$ rather than $\bm{\hat{E}}=-\partial_t\bm{\hat{A}}$, but the calculation goes through as before\cite{merzbacher} and thus a GCS can be generated by acting on a number state with a classical external current. This fact was noted by Oliveira {\it et al.},\cite{oli90} who point out that the preparation of number states can be carried out in a cavity, which can then be driven by a current that is classical up to negligible quantum fluctuations. It appears however that quantum-optical GCS have yet to be generated experimentally, though GCS of a quantum oscillator consisting of a trapped ion have very recently been reported.\cite{zie13}

\section{Conclusions}
We have given a basic description of generalized coherent states (GCS) in quantum optics. These states are an interesting and natural accompaniment to presentations of number states and coherent states. GCS combine features of number states and coherent states in a nontrivial manner. The electric-field probability distributions of GCS show the phase oscillation of coherent states but also contain nodes that behave like single photons. This allows GCS to have macroscopic energies while still having some of the properties of few-photon number states. 

\begin{acknowledgments}
I thank S. Horsley for helpful discussions. I am also indebted to E. K. Irish for information on literature.
\end{acknowledgments}

\end{document}